\documentclass[jcp,twocolumn,showpacs,preprintnumbers,amsmath,amssymb]{revtex4}
\usepackage{graphicx}
\usepackage{dcolumn}
\usepackage{bm}
\newcommand{\be}[1]{\begin{equation}\label{#1}}  
\newcommand{\ee}{\end{equation}}  
  
\begin{document}

\title{Femtosecond Photoionization of Atoms under Noise}
\author{Kamal P. Singh and Jan M.~Rost}  
\affiliation{Max Planck Institute for the Physics of Complex Systems, N\"othnitzer Strasse 38, 01187 Dresden, Germany. }

\begin{abstract}  

  We investigate the effect of incoherent perturbations on atomic 
  photoionization due to a femtosecond mid-infrared laser pulse by
  solving the time-dependent stochastic Schr\"odinger equation. 
  For a weak laser pulse which causes almost no ionization, 
  an addition of a Gaussian white noise to the pulse leads to a 
  significantly enhanced ionization probability. 
  Tuning the noise level, a stochastic resonance-like curve
  is observed showing the existence of an optimum noise for a given 
  laser pulse. Besides studying the sensitivity of the obtained enhancement
  curve on the pulse parameters, such as the pulse duration and peak
  amplitude, we suggest that experimentally realizable broadband 
  chaotic light can also be used instead of the white noise to 
  observe similar features. 
  The underlying enhancement mechanism is analyzed in the frequency-domain 
  by computing a frequency-resolved atomic gain profile, 
  as well as in the time-domain by controlling the relative delay between 
  the action of the laser pulse and noise.  

\end{abstract} 

\maketitle  
 
\section{INTRODUCTION}  

 The role of noise on driven quantum systems is a subject of considerable
 interest and importance. 
 Several studies exist in the literature on this broad topic. For instance, 
 the noise-induced effects in the nanoscale quantum devices such as
 Josephson junctions \cite{Gamat, Buchl}, the macroscopic phase transitions due to quantum
 fluctuations \cite{Griener}, driven multilevel quantum systems under incoherent 
 environment \cite{RabitzReview}, and stochastic ionization of Rydberg atoms by 
 microwave noise \cite{NoiseIonRyd}, to mention just a few.
 In these examples, the action of noise on a system can be broadly classified as 
 being of two types: either of destructive nature, i.e., noise must be
 avoided; or of constructive nature necessitating its presence. 

 It is this nontrivial latter aspect of noise-induced effects that
 has been subject of intense investigation \cite{Gamat, Buchl, Badzey, Moss}.
 In particular, the stochastic resonance phenomenon (SR)
 provides a paradigm for the constructive role of noise in nonlinear
 classical as well as quantum systems \cite{Moss}.
 The essence of quantum SR is the existence of an optimum amount of 
 noise in a nonlinear system that enhances its response
 to a weak coherent input forcing \cite{QSR, mMaser, Huelga, H2O}.
 Despite the diversity of nonlinear dynamics exploiting classical SR, 
 most of the quantum mechanical studies of the effect have focused on
 the so-called spin-Boson model, which provides an analog of
 a classical double-well potential \cite{Gamat, Buchl, Makri, Grifoni}.
 However, many other physical systems exist, particularly 
 atoms or molecules exposed to strong laser pulses, where the 
 quantum dynamics can be nonlinear and therefore added noise
 could play an important role. 

 The presence of noise is also worth studying from 
 the point of view of steering quantum dynamics of atomic or molecular
 systems \cite{Rabitz, Zewail, Assion}. Traditionally this is achieved with strong,
 tailored laser pulses by exploiting their nonperturbative
 and nonlinear interaction with the atomic systems. In this context,
 it has been shown that weak noise of various origin
 in multilevel ladder systems plays a crucial role \cite{Rabitz}. 
 Many scenarios have been discussed, such as the need to either
 cooperate or fight with de-coherence in the closed-loop control \cite{Rabitz-1}, 
 and to engineer the environment to achieve
 the steering of quantum systems towards a desired state \cite{Rabitz-2}.
 In the same spirit, white shot noise has been used to dissociate
 diatomic molecules \cite{kenny}. This has 
 implications for the field of quantum control.
 Motivated by the concept of exploiting noise in nonlinear systems,
 one can ask the question if noise can serve as an extra tool for quantum control. 
 Indeed, the concept of the quantum SR effect has not been 
 exploited for additional insight in controlling the quantum phenomenon. 
 
 In this article, we provide a detailed study of the influence of noise
 in a generic quantum situation, namely the photoionization of 
 a single-electron atom interacting with an ultrashort laser pulse. 
 We will demonstrate the conditions under which a
 resonance-like behavior emerges in the stochastic photoionization
 process. This noise-induced phenomenon is studied for a variety of 
 laser pulses, from a few optical cycles duration to very long ones, 
 and of varying intensities.
 Furthermore, we suggest the experimental observability of the effect
 by employing a broadband chaotic light instead of white 
 Gaussian noise. Lastly, we characterize the underlying gain-mechanism
 in the frequency-domain in order to identify the 
 crucial frequency-bands in the broad noise spectrum.

 The article is organized as follows. Section II introduces our 
 model of the simplest atom interacting with a femtosecond laser pulse 
 and white noise, and describes our method to solve its 
 stochastic Schrodinger equation.
 In section III, we show the results of the ionization probability
 (IP) for various combinations of the laser pulse and noise.
 The existence of a stochastic resonance-like behavior is quantified
 using an enhancement factor which is computed from IP.
 The sensitivity of this noise-induced effect is tested with
 laser pulses of varying duration and strength,
 and other types of noise such as a realizable broadband chaotic light.
 To characterize the enhancement mechanism, we compute
 the frequency-resolved gain profile of the driven atom, and 
 study the role of relative time-delay between noise and
 the laser pulse. Finally, section IV provides a summary of 
 results with our conclusions.

\section{DESCRIPTION OF THE MODEL}  

\subsection{Hydrogen atom interacting with a laser pulse and noise}  

 We consider as an example the simplest single-electron atom, 
 i.e., the hydrogen atom. Due to the application of an intense 
 linearly polarized laser field $F(t)$, the electron dynamics is 
 effectively confined in one-dimension along the laser polarization axis \cite{Eberly}.
 The Hamiltonian for such a simplified description of the hydrogen atom,
 which is here also perturbed by a stochastic force $\xi(t)$ \cite{kamal}, 
 reads as (atomic units, $\hbar=m=e=1$, are used unless stated otherwise),
 \begin{equation}  
   H(x,t) = \frac{\hat{p}^2}{2} + V(x) + x \lbrace F(t) + \xi(t) \rbrace,
   \label{eqn:eqn1}
 \end{equation}
 where $x$ is the position of the electron and \mbox{$\hat{p}=-i\;\partial/\partial x $} 
 is the momentum operator. The external perturbations, $F(t)$ and $\xi(t)$, 
 are dipole-coupled to the atom. The potential is approximated 
 by a non-singular Coulomb-like form,
 \begin{equation}  
     V(x)=- \frac{1}{\sqrt{x^2 + a^2}}. 
   \label{eqn:eqn2}
 \end{equation}
 Such a soft-core potential with parameter $a$ has been routinely employed
 to study atomic dynamics in strong laser fields \cite{mpi}. It successfully
 describes many experimental features of multiphoton 
 or tunnel ionization \cite{Eberly}, and the observation of the plateau in
 higher harmonic generation spectra \cite{mpi}. 

 The laser field is a nonresonant mid-infrared (MIR) femtosecond pulse 
 described as, 
 \begin{equation}  
 \mbox{$ F(t) = f(t) F_0\sin(\omega t+\delta)$}.
 \label{Lsrpulse}
 \end{equation}  
 Here $F_0$ defines the peak amplitude of the pulse, $\omega$ denotes
 the angular frequency, and $\delta$ is the carrier-envelop phase. 
 We choose a smooth pulse envelop $f(t)$ of the form,
 \begin{displaymath}  
 f(t) = \left\{ \begin{array}{ll}
      \sin^{2}(\pi t/(2 \tau)),    & \textrm{$t < \tau$} \\
      1,                        & \textrm{$\tau \leq t \leq T_p-\tau$} \\
      \cos^{2}(\pi (t + \tau - T_p)/(2 \tau)), & \textrm{$T_p-\tau < t \leq T_p$},
      \end{array} \right.
    \label{eqn:envelop}
 \end{displaymath}
 where $T_p$ is the pulse duration and $\tau$ the time for turning the field on
 and off.

 The noise term $\xi(t)$ is a zero-mean
 Gaussian white noise having the following properties,
 \begin{equation}   
   \langle \xi(t) \rangle = 0,
   \label{eqn:eqn2a}
 \end{equation}    
 \begin{equation}   
   \langle \xi(t)\xi(t')\rangle = 2D\;\delta (t-t'),
   \label{eqn:eqn2b}
 \end{equation}                  
 and noise intensity D \cite{book}.

 \begin{figure}[t]   
    \includegraphics[width=.95\columnwidth]{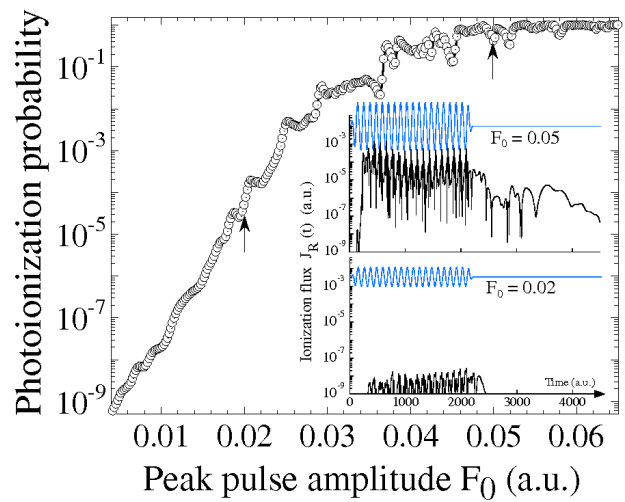}
 \caption{
 The ionization probability $P_l$ as a function of the laser peak 
 amplitude $F_0$. Insets show the ionization flux versus time
 for two different pulses (top parts of curves) of amplitudes,
 $F_0=0.05$ and $F_0=0.02$, marked by arrows on IP curve.
 Here $\omega=0.057$, $\delta=0.0$, $T_p=20\pi/\omega$
 and $\tau=2 \pi/\omega$.
  }
\end{figure}

\subsection{Stochastic quantum dynamics} 

 The presence of the stochastic forcing term in the Hamiltonian
 as described above, makes the quantum evolution nondeterministic. 
 Thus an averaging over a large number of realizations of the stochastic force
 is required in order to produce a statistically meaningful solution of the 
 following time-dependent stochastic Schr\"odinger equation,
 \begin{equation}    
   i\; \frac{\partial \Psi(x,t)}{\partial t} = H(x,t)\; \Psi(x,t).
 \label{eqn:eqn6}
 \end{equation}
 For a given realization, the numerical solution of the  Schr\"odinger
 equation amounts to propagating the initial wave function 
 $\vert \Psi_0 \rangle$ using the infinitesimal short-time stochastic
 propagator,
 \begin{equation}   
  U_\xi (\Delta t) = \exp\left( -i\int_{t}^{t+\Delta t} H(x,t) dt\right). 
 \label{eqn:eqn2a}
 \end{equation}  
 One can compute $U_\xi (\Delta t)$ using the split-operator fast Fourier
 algorithm \cite{SOper}. Details of the  method employed are described in
 the Appendix. Successive applications of the stochastic propagator 
 $U_\xi (\Delta t)$ advance $\vert \Psi_0
 \rangle$ forward in time. 

 Note that the initial state $\vert \Psi_0 \rangle$
 is always chosen to be the \emph{ground state} of the system 
 having an energy of $I_b = -0.5$~a.u..
 This is obtained by the imaginary-time relaxation
 method for $a^2=2$ \cite{Eberly}.  To avoid parasitic reflections of the 
 wavefunction from the grid boundary, we employ an absorbing boundary \cite{SOper}. 
 
 The ionization flux leaking in the continuum on one side, 
 is defined as \cite{QMbook},
 \begin{equation}   
  J_R(x_R,t) = Re\lbrack \Psi^\ast \: \hat{p} \: \Psi \rbrack_{x_R} ,
 \label{eqn:Jflux}
 \end{equation}  
 where $x_R$ is a distant point (typically 500 a.u.) near the absorbing
 boundary. The ionization rate is integrated over a sufficiently long time
 interval to obtain the corresponding total ionization probability,
 \begin{equation}  
  P = \int_{0}^{\infty} J_R(x_R,t) dt.
 \label{eqn:IP}
 \end{equation}  

 In the following section, we shall use both the ionization flux, and the
 photoionization yield, to study the interplay between the laser pulse 
 and noise. From the point of view of stochastic resonance-like
 phenomena, we aim at establishing the constructive role of noise
 in atomic photoionization due to a femtosecond laser pulse.

\section{RESULTS}

\subsection{Optimal stochastic enhancement of photoionization}  

\subsubsection{Photoionization as a nonlinear effect}  

 Let us first consider the response of the atom interacting with
 a short but strong laser pulse only. Fig.~1 shows the ionization 
 probability $P_l$ versus the peak pulse amplitude $F_0$ for a 
 20 cycle long MIR laser pulse ($ \omega=0.057$). This figure shows 
 that with increasing values of $F_0$ the ionization probability 
 first increases nonlinearly, and then saturates to the maximum value of
 unity, for $F_0>0.05$. The behavior of $P_l(F_0)$ is a characteristic 
 signature for many atomic and molecular systems interacting 
 with nonresonant intense laser pulses \cite{mpi}.

 The laser pulse produces (nonlinear) ionization of the atom which
 is most easily understood, especially in the time domain,
 with the picture of a periodically changing tunneling barrier. 
 Ionization flux is produced close
 to those times when the effective potential \mbox{$U(x,t)=V(x)+x F(t)$},
 is maximally bent down  by the dipole-coupled laser field.
 This is illustrated in the inset of Fig.~1 with the temporal evolution of the
 ionization flux  for laser pulses (shown in the top parts of
 inset) with two different peak amplitudes $F_0=0.05$ and
 $F_0=0.02$ (see arrows in Fig.~1).
 Time-resolved ionization peaks separated by the optical period ($2\pi
 / \omega$) are clearly visible for both peak field amplitudes. 
 In addition, $J_{R}(t)$ shows a complex interference pattern [inset of Fig.~1] due to
 the modulated Coulomb barrier for $F_0=0.05$. However, quite strikingly, 
 if  $F_0$ is reduced to 0.02 a.u., 
 the ionization flux collapses by around five orders of magnitude as shown
 in Fig.~1. One can therefore conclude that the photoionization dynamics is highly
 nonlinear, and in particular it exhibits a form of ``threshold''
 dynamics where the threshold is created by the condition 
 for over-the-barrier ionization. 

\begin{figure}[t]
  \begin{center}
    \includegraphics[width=.90\columnwidth]{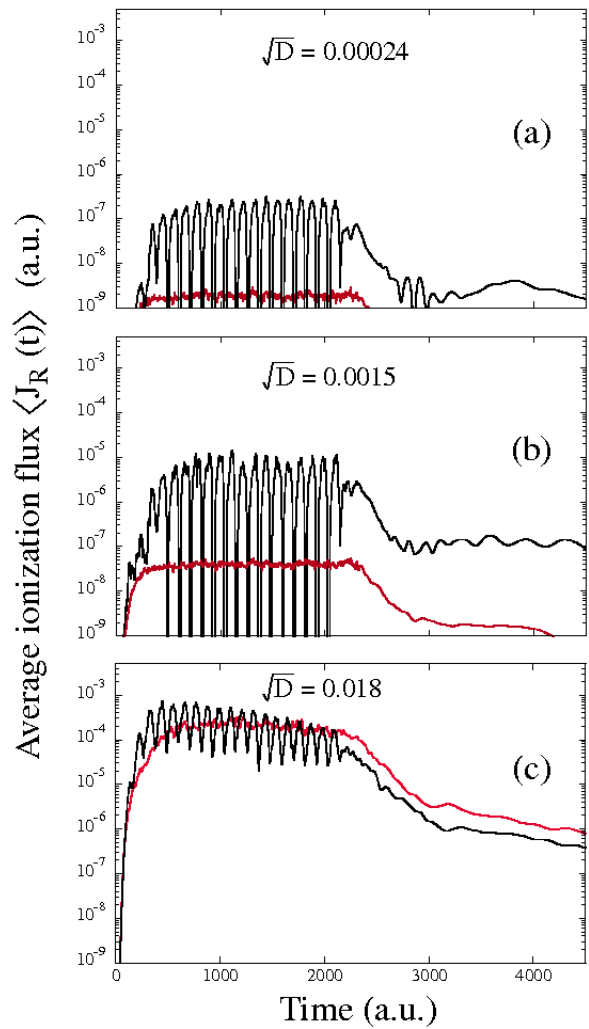}
    \caption{
 Ionization flux for a weak laser pulse $F_0=0.02$,
 with three values of noise amplitude, (a) $\sqrt{D}=0.00024$, 
 (b) $0.0015$, and (c) $0.018$. Background featureless curves (red)
 show the corresponding purely noise-driven ($F_0=0$) flux.
 The flux is averaged over 50 realizations.}
\label{fig:fig2}
   \end{center}
\end{figure}

 \subsubsection{Ionization induced by noise alone}  

 Here we look into the possibility of efficiently ionizing the atom, when
 it is subjected to white Gaussian noise only. The interaction time of
 the atom with the noise is kept identical to the laser pulse duration
 $T_p$ (see inset of Fig.~3). Fig.~2 shows the evolution of the 
 ionization flux $\langle J_R(t) \rangle$
 which is averaged over 50 different realizations of the noise. 
 One can see that for small noise amplitudes the ionization flux exhibits
 a featureless curve, producing the ionization flux around $10^{-9}$. 
 As the noise level is increased, the featureless ionization curve
 rises monotonically as shown in Fig. 2(a)-(c). 
 By integrating the stochastic ionization flux, one can compute the 
 corresponding ionization probability $P_n$, which is simply equal to
 the area under the curve $\langle J_R(t) \rangle$. 
 The resulting noise-induced ionization probability $P_n$ versus the noise
 amplitude is shown in Fig.~3. As can be seen the stochastic ionization 
 probability rises monotonically with the noise level.
 For ultra-intense noise, such that its strength becomes comparable to the 
 atomic binding field, obviously full ionization can be achieved.
 We should mention that similar effects have been observed in 
 other systems, for example, the purely noise-induced molecular
 dissociation \cite{kenny}, and the ionization induced by weak noise 
 of the highly excited Rydberg atoms \cite{NoiseIonRyd}.
 However, in our case we consider the atom to be initially in its ground state.

 \begin{figure}[t]    
 \includegraphics[width=.9\columnwidth]{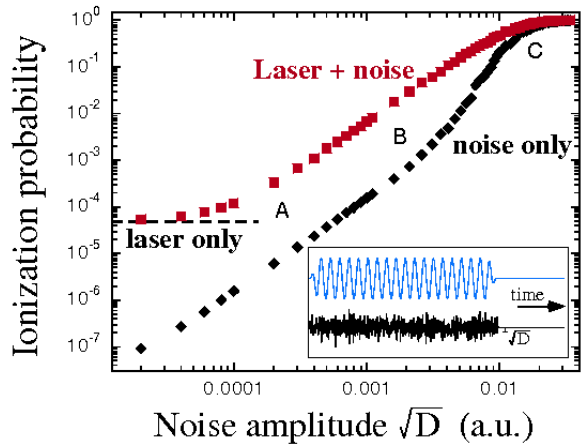}
 \caption{The ionization probability versus the noise amplitude $\sqrt D$ for the
 noise alone $P_n$ (diamonds) and for the laser pulse ($F_0=0.02$)
 with noise $P_{l+n}$ (squares). The points A-C corresponds to the three
 cases considered in Fig.~2. The dashed line shows the limit of vanishing
 noise, i.e., the probability due to the laser pulse alone $P_l$. 
 Inset: an example of laser pulse and noise
 }
 \label{fig:fig3}
 \end{figure}

 Although the application of noise alone, or the laser pulse alone, can
 lead to the atomic ionization, we aim to study whether a combination
 of both the laser pulse and noise makes the ionization process more
 efficient as compared to the individual cases.

 \subsubsection{Simultaneous application of the laser pulse and noise}  

 We have seen that the atomic photoionization due to an intense
 femtosecond laser pulse is a highly nonlinear quantum phenomenon, 
 and in particular, the ionization response collapses for a ``weak''
 laser pulse (see inset of Fig.~1).
 Motivated by the quantum SR effect, we wish to explore
 if the noise can recover the strong periodic ionization flux for the weak
 laser pulse. To answer this question, in Fig. 2(a) we show the average 
 ionization flux when a small noise
 of amplitude, $\sqrt{D} =0.00024$, is added to
 the previously weak laser pulse ($F_0=0.02$). 
 Note that the atomic excitation time by the
 laser and the noise here are identical. One can see that for such a feeble noise
 amplitude, the periodic structure in atomic ionization gets enhanced by
 more than one order of magnitude, as compared to the case of the noise alone
 which is shown as the background featureless curve.
 Hence, the observed net enhancement can be attributed 
 to a nonlinear quantum interaction between the coherent pulse and noise.

 \begin{figure}[t]    
 \includegraphics[width=.9\columnwidth]{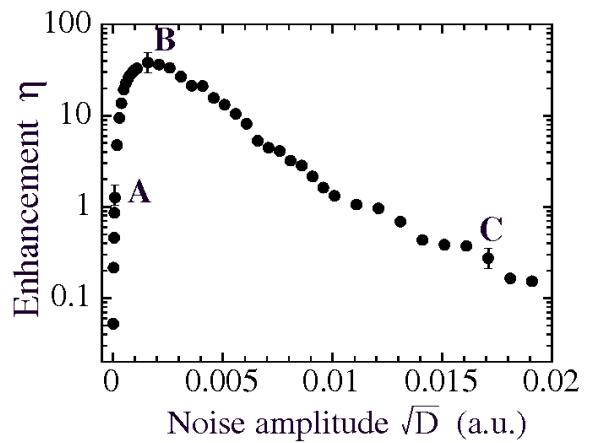}
 \caption{The enhancement in photoionization due to quantum SR.
 The points marked A-C correspond to the  noise amplitudes of
 Fig.~2(a)-(c), respectively. At point B the ratio
 $\sqrt{D_{opt}}/F_0 = 0.075$. The error bars indicate the
 standard deviation of $\eta$ calculated using more than 1000
 different noise realizations.
 }
 \label{fig:fig4}
 \end{figure}

 As the noise level is further increased, we observe an enhancement of 
 the periodic ionization profile by around three orders of magnitude 
 as shown in Fig.~2(b). However, the increase in noise level also 
 causes the background structureless stochastic ionization curve to rise
 monotonically. For strong noise (Fig. 2(c)), these periodic structures 
 tend to wash out and the process is effectively controlled by the noise.
 Hence one expects, the existence of an intermediate noise level 
 where the nonlinear ionization is optimally enhanced.

 The net enhancement of the atomic ionization due to interplay
 between the laser pulse and the noise can be characterized by
 the enhancement factor \cite{kamal},
 \begin{equation}            
 \eta = \frac{P_{l+n} - P_0}{P_0},
 \label{eqn:eqn4}
 \end{equation}
 with $P_{0} = P_l + P_n $. Although this is different compared to the
 quantifiers commonly used \cite{Gamat,Buchl}, $\eta$ is more suitable for our case.
 One can verify that a zero value of $\eta$ corresponds to the case 
 when either the laser pulse ($P_l \gg P_n$) or the noise ($P_l \ll P_n$) dominates.
 In Fig.~4, we have plotted the enhancement factor
 $\eta$ versus the noise amplitude $\sqrt{D}$.
 It exhibits a sharp rise, followed by a maximum at a certain value of the
 noise (point B), and then a gradual fall off. 
 It is worth mentioning that only a modest noise-to-laser ratio 
 ($\sqrt{D_{opt}}/F_0 = 0.075$) is required to reach the optimum enhancement 
 (here $\eta_{max} = 36$). 

 Before investigating other properties of the enhancement effect,
 it is worth making three remarks.
 First, although the enhancement curve bears striking resemblance 
 to the typical SR curve, it is not SR effect where the matching of
 the time scales between coherent and incoherent driving exists. 
 The underlying gain mechanism here is completely different,
 as we shall see later. One can perhaps call
 this as a generalized quantum SR for such atomic systems,
 in the sense of the existence of an optimum noise level.
 Second, the location of the optimum enhancement 
 is governed by an empirical condition, when the strengths of the laser
 pulse and noise are comparable, in terms of the ionization flux produced by
 their individual action $P_l \sim P_n $. This can be verified in Fig.~3 for the 
 enhancement curve shown in Fig.~4.
 Third, due to the presence of the random noise term in the Hamiltonian, 
 the optimal solution is only statistically unique. We have computed the standard
 deviation of the enhancement factor $\eta$ using 1000 realizations,
 $\sigma = \sqrt{<\eta^2> - <\eta>^2}$. The corresponding $\sigma$ values 
 are shown by error bars on the $\eta$ curve in Fig.~4.

 \subsubsection{Enhancement curves for a variety of laser pulses}  

 In this subsection, we study the sensitivity of the stochastic enhancement 
 curves on the laser pulse. In particular, we study the role
 of two parameters: (i) the pulse duration $T_p$, 
 and (ii) the peak pulse amplitude $F_0$. 
 In Fig.~5, we have plotted enhancement curves versus the noise amplitude
 for pulses of fixed amplitude ($F_0=0.02$) but of varying durations
 from 5 to 30 optical cycles. One can clearly
 see that, the enhancement features (particularly the location and
 strength of the optima) are robust for pulses
 ranging from ultrashort few cycle duration to quite long ones. 

 Although we do not show, we have also verified that variation in the carrier
 envelop phase $\delta$ of the laser pulse $F(t)$ [see Eq.~(3)] does not
 modify the enhancement effect. 
 This can be expected due to the presence of the noise term, 
 by which any effect of $\delta$ is averaged out.
 Furthermore, we have also observed
 similar enhancement curves for other forms of the 
 pulse envelop, such as $f(t)=\sin^{2}(\pi t/T_p)$. 

 \begin{figure}[t]    
 \includegraphics[width=.95\columnwidth]{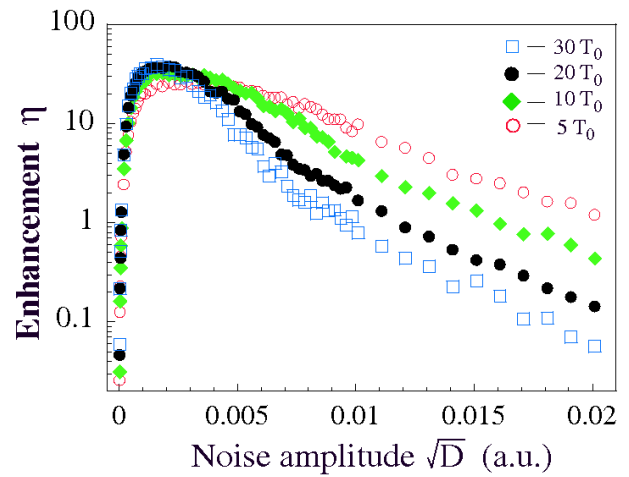}
 \caption{The enhancement factor versus the noise amplitude 
 for four different values of pulse durations
 $T_p$: 5, 10, 20, 30 optical period $T_0$. Here $F_0=0.02$, and other
 parameters are the same as in previous figures.
 }
 \label{fig:fig5}
 \end{figure}

 To investigate the dependence of $\eta$ on the laser pulse amplitude $F_0$,
 we choose some moderate noise amplitude value, for example,
 $\sqrt{D}=0.0015$. For this fixed $\sqrt{D}$, we now increase 
 the peak pulse amplitude $F_0$ of the 20 cycles pulse
 from zero to a large value, and calculate the IP for each value of $F_0$. 
 The obtained probabilities for different cases are plotted in Fig.~6(a),
 which are then used to compute the enhancement factor $\eta$ in Fig.~6(b). 
 Here, $\eta$ also exhibits a nonmonotonic feature versus the laser peak
 amplitude, thus suggesting a range of 
 $F_0$ where the addition of noise can be useful. 
 This dependency is intuitively explained. Since for weak laser pulse  
 the process is dominated by the noise and the $\eta$ collapses.
 On the other hand, if the laser peak amplitude is too strong (comparable to
 over-the-barrier ionization threshold), the pulse can ionize
 the atom by itself, and the noise has no role to play.
 Thus, it is for the intermediate values of the noise and laser
 pulse amplitudes, where this nonlinear enhancement mechanism can be significant. 
 From Figs.~4 and~6, one can conclude that
 in order to maximize the net ionization yield,
 a particular pair of $F_0$ and $\sqrt{D}$ is required.

 \begin{figure}[t]    
 \includegraphics[width=.9\columnwidth]{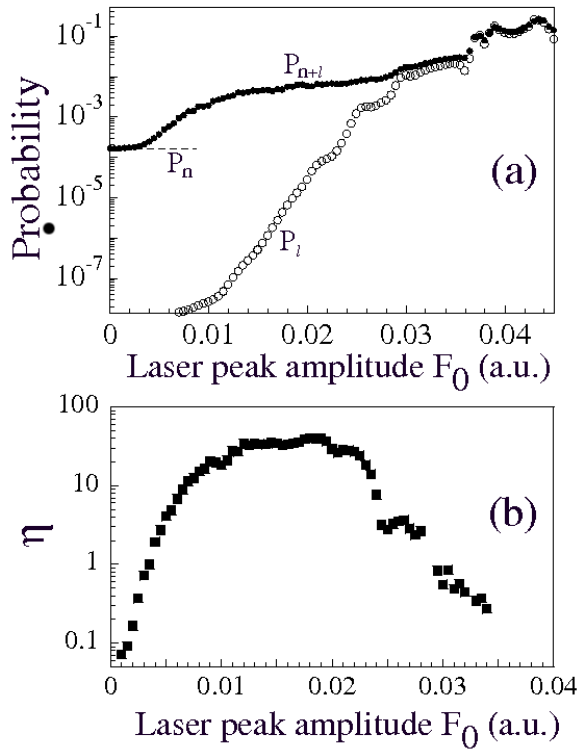}
 \caption{ The dependence of enhancement on $F_0$ for a 
  fixed noise amplitude $\sqrt D=0.0015$. (a) Ionization probabilities
  versus $F_0$ for laser only $P_l$, and for laser with noise $P_{l+n}$. 
 The dashed line shows the limiting value of 
 $P_n$. (b) The enhancement vs $F_0$ curve obtained from the probability
 curves shown in (a).
 }
 \label{fig:fig6}
 \end{figure}

 \subsection{Employing chaotic light instead of the white noise}  

 \subsubsection{Generation and characterization of chaotic light} 

 To experimentally observe this effect, the most challenging
 task is the generation of intense white noise. 
 For instance, if one considers the thermal radiation from a
 blackbody (such as the sun) as a possible source of
 the white noise, its noise intensity falls short by many
 orders of magnitude \cite{Dunbar},
 compared to the one required for the optimum of Fig.~4.
 We thus look into alternatives 
 for generating a noise-like waveform. One possibility is
 to employ modern pulse shaping techniques,
 whereby one can design waveforms of almost 
 arbitrary shapes \cite{chaoLgt, MIRshaping}.
 To realize such a chaotic light, we choose a large number of 
 frequency modes, $N$, in a finite but broad bandwidth $\Delta\omega$. 
 These modes can be, for example, different Fourier components of an 
 ultrashort laser pulse. The total electric field $Z(t)$ is a sum
 of these $N$ individual modes as  \cite{Weiner},
 \begin{equation}            
 Z(t) = \sqrt{\frac{2}{N}} \sum_{n=1}^N F_{rms}\sin(\omega_n t + \phi_n), 
 \label{eqn:eqn4}
 \end{equation}
 where $\omega_n$, $\phi_n$ denote the angular frequency, phase of 
 $n^{th}$ mode, respectively; and $F_{rms}$ is 
 the root-mean-square amplitude of $Z(t)$.
 Note that here we consider these frequency modes to
 oscillate independently with their phases $\phi_n$
 assuming random values relative to each other. 
 In this particular case of phase-randomized coherent modes, 
 the total field $Z(t)$ at any point will be noise-like, fluctuating
 in intensity due to the interference between modes. 
 Inset in Fig.~7 shows an example of such a chaotic
 light spectrum for $N=1024$ in a chosen bandwidth (BW) of 0.75
 (corresponding to a 32 attosecond pulse) \cite{Martn}. Such a construct 
 tends to the white noise, in the limit of $\Delta\omega, N \to \infty $.
 In the following subsection, we consider a simultaneous application of the 
 weak laser pulse and chaotic light (instead of the white noise) and
 see if the enhancement effect can be preserved. 

 \subsubsection{Photoionization by the chaotic light} 

 When we replace white noise by the chaotic light (BW=0.75 and $N=1024$),
 one can see in Fig.~7 that most of the features of the enhancement
 phenomenon remain intact. In particular, the intensity and location of the 
 optimum is very close to the one obtained for the white noise case in Fig.~4.
 We also recover other features of the enhancement mechanism with the chaotic light
 such as its dependence on the pulse duration $T_p$ and pulse amplitude 
 $F_0$.
 
 This observation not only suggests the possibility
 of observing the effect using a finite but broadband chaotic light,
 but also raises the question concerning the relevant frequency components in
 the chaotic light spectrum, to be discussed next.

 \begin{figure}[t]    
 \includegraphics[width=.9\columnwidth]{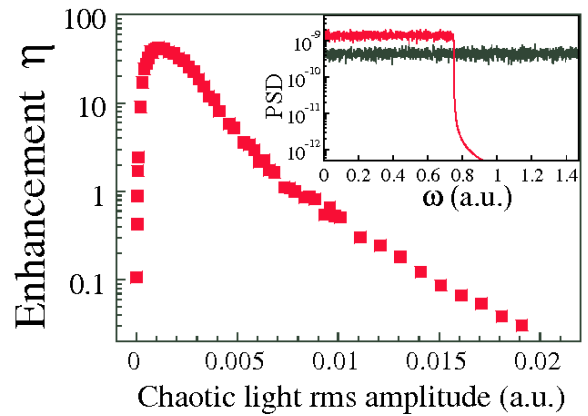}
 \caption{Enhancement induced by a broadband chaotic light. 
 The peak amplitude and frequency of the 20 cycle laser pulse are 
 $F_0=0.02$ and $\omega=0.057$, respectively. The bandwidth of chaotic light
 is $\Delta\omega=0.75$ with central frequency $\omega_0 =0.375$.
 Inset: power spectral density (PSD) of the chaotic light 
 compared to the one for the white noise.
 }
 \label{fig:fig7}
 \end{figure}

 \subsection{Spectral and temporal analysis of the gain mechanism}  

 \subsubsection{Frequency-resolved gain profile} 

 In this subsection, we will analyze the mechanism of stochastic 
 enhancement in both, the frequency domain and time domain. 
 In particular, we aim to identify frequency components
 in the broad spectrum of noise (or chaotic light) which
 are the crucial ones to provide the gain. 
 For this purpose, we compute a frequency-resolved 
 atomic gain (FRAG) profile using a pump-probe type of
 setting as described below.

 \begin{figure}[t]    
 \includegraphics[width=.9\columnwidth]{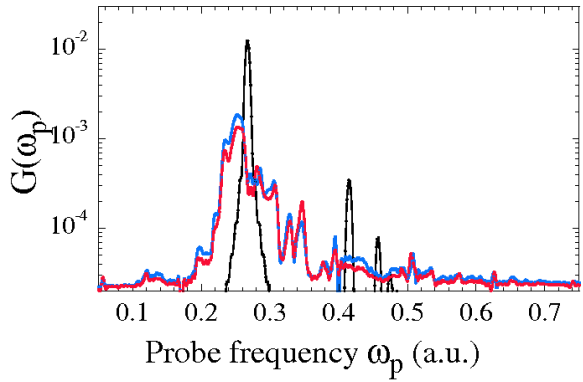}
 \caption{Frequency-resolved gain $G(\omega_p)$ of the atom 
   which is driven by a 10 cycle laser pulse of $F_0=0.02$ at 
  $\omega = 0.057$. The atomic gain is probed by a simultaneous
  application of a weak probe beam $F_p = 0.0002$ of tunable 
  frequency $\omega_p$. Both, the depletion of the ground state 
  population (blue) and the net absorption of energy (red) due to the probe
  are plotted. The gain of driven atom is also compared
  with the bare atomic gain (black).
 }
 \label{fig:fig8}
 \end{figure}

 It is well known that when an atom interacts linearly with a weak 
 external field, the energy absorption takes place at its
 resonant frequencies. However, due to its interaction with a
 strong laser pulse, the unperturbed atomic states are 
 significantly modified leading to a completely different frequency 
 response of the driven atom. Indeed, such a modified spectral response
 is the relevant quantity when the atom is also subjected to noise.
 So, how can one measure precisely the frequency resolved gain
 $G(\nu)$ offered by the atom? One possible way to observe FRAG 
 is to consider the previously employed laser pulse
 ($\omega=0.057$, $F_0 = 0.02$) as a pump pulse
 and replace noise by a tunable monochromatic probe pulse,
 \mbox{$ F_p(t) = f(t) F_p \sin(\omega_p t)$}. The probe amplitude $F_p$ 
 is much weaker than the driving laser pulse [$F_p / F_0 = 0.01$],
 such that it does not significantly
 alter the quasi energy levels, but can drive transitions between them.
 For an atom prepared in its ground state, a FRAG profile is obtained 
 by measuring either the depletion of the ground state population
 or the net energy absorption, as a function
 of the probe frequency $\omega_p$.
 
 Such a gain profile $G(\omega_p)$ is shown in Fig.~8 for the
 atom driven by a 10 cycle long laser pulse of amplitude $F_0=0.02$.
 Although the FRAG shows lots of structure,
 the dominant peaks appear around the first atomic transition frequency.
 A closer inspection of Fig.~8 reveals that there is a dip at 
 the unperturbed resonance frequency $\omega_{01}=0.267$.
 The main peak is indeed shifted beyond the linear Stark shift for our model.
 These shifts and broadening of the atomic resonances are caused by
 the strong laser pulse, since the field can drive the electron 
 significantly away from the nucleus leading to strong nonlinear
 perturbation to its bound states. 
 Such a FRAG curve provides a fingerprint of the atomic gain under the strong
 laser pulse. The frequency bands that correspond to the peaks in 
 the obtained gain curve are indeed the most useful ones to obtain the enhancement.
 In the following, we will test the validity of this statement by designing a
 chaotic light where significant resonance frequencies are filtered from
 its spectrum.

 \subsubsection{Chaotic light with missing resonant frequencies} 

 To show that the useful frequencies are \emph{not} simply the resonant 
 frequencies of the atom, we have designed a chaotic light spectrum
 perforated by digging holes in its spectral density
 around the first few atomic transition frequencies. 
 As can be seen from Fig.~9, almost no enhancement is lost,
 if the hole width $w$ is below 0.013~a.u., 
 which already includes the linear ac stark shift of the atomic states. 
 However, by increasing the hole width such that no frequency component
 exists in the noise spectrum where the FRAG has dominant peaks 
 leads to a collapse of the enhancement mechanism, as shown in Fig.~9. 
 This observation validates the importance of FRAG structure in 
 identifying the useful spectral bands in noise. It suggests that
 for the simultaneous presence of the laser pulse and noise,
 non-resonant frequency components are the dominant ones. 

 \begin{figure}[t]    
 \includegraphics[width=.9\columnwidth]{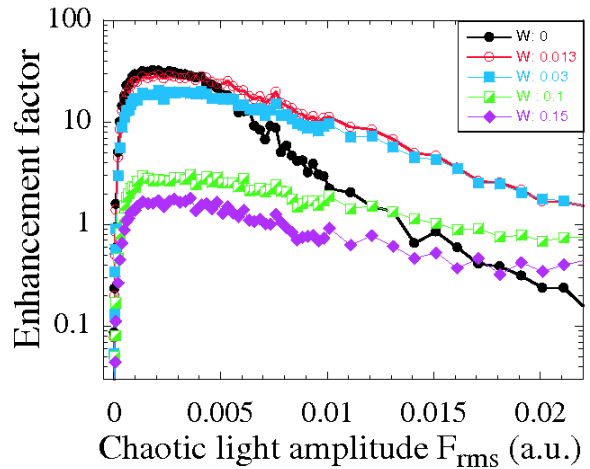}
 \caption{Enhancement curves due to chaotic light spectrum
 perforated by holes at the first three resonance frequencies. 
 Difference curves correspond to the increasing hole widths
 of $w=0.0$, 0.013, 0.03, 0.1 and 0.15.
 }
 \label{fig:fig9}
 \end{figure}

 It is worth making few remarks here. First, the FRAG is a property of
 the atom interacting with a particular strong laser pulse. Thus, the
 detailed features of the gain curve depends on both the atomic system
 and on the laser pulse parameters. But this doesn't affect the general
 conclusion drawn from such a curve.
 Second, it is also possible to obtain the enhancement using a 
 monochromatic beam instead of the chaotic broadband light, 
 if its frequency is properly tuned to the new ``resonances''. But, 
 the advantage of using a broadband source is that the enhancement becomes 
 independent of both, the particular atom and the FRAG structure
 for different pulse parameters.

 \subsubsection{Role of relative delay between signal and noise} 

 \begin{figure}[t]    
 \includegraphics[width=.9\columnwidth]{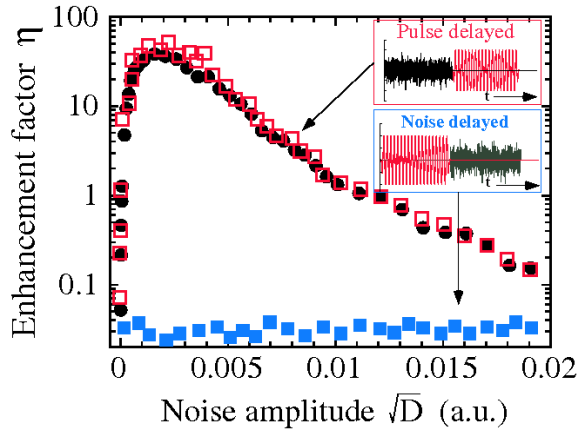}
 \caption{Enhancement factor $\eta$ versus the noise amplitude $\sqrt D$
 for various cases, application of noise first and then the laser pulse
 (empty squares), application of laser pulse first and then noise
 (filled square), simultaneous application of laser and noise (circles).
 Insets depicts schematically temporal signal applied to the atom
 for corresponding cases.  
 }
 \label{fig:fig10}
 \end{figure}

 Up to now we have considered the case of a perfect synchronization, i.e., 
 a simultaneous application of the laser pulse and noise
 to observe the enhancement. 
 We now wish to relax this synchronization constrain between the 
 laser pulse and noise to test if the enhancement still exists. 
 Such a scenario would not only help the experimental search of similar
 effects, but also provides an alternative aspect of the enhancement mechanism as 
 compared to the one mentioned above.

 There are two possible ways to expose the atom to a laser pulse 
 and noise sequentially:
 (i) the atom first interacts with noise only and then we apply the laser pulse,
  or conversely, (ii) the laser pulse is applied first and then the noise is applied. 
 Note that for both cases, there is no direct interplay between
 laser and noise. The enhancement factor $\eta$, which is defined as before, 
 is shown in Fig.~10 for both the cases. 
 One can clearly see that for the case-(i) $\eta$ curve looks very similar 
 to the  one for simultaneous action of laser and noise. But 
 for the case-(ii), the enhancement curve collapses.

 Although we obtain almost identical enhancement curves for cases of,
 first-noise then laser pulse and simultaneous action of noise with laser pulse,
 the gain providing frequency components are fundamentally different
 in each case. The gain curves for both cases are given in Fig.~8.
 For the sequential application of noise and laser,
 the atomic gain is basically at the resonant frequencies of the unperturbed atom.
 Thus the resonant frequencies (particularly the first few)
 are the most fertile ones in the noise spectrum.
 In this case, a two-step picture of the enhancement mechanism applies, where
 the atom first absorbs energy from the noise leading to an
 exponential population distribution, and the laser causes ionization 
 from ``noise-heated'' atom in a second step.

\section{Summary and Conclusion}  

 We have investigated photoionization of a hydrogen atom
 which is subjected to both, a MIR femtosecond laser pulse and white Gaussian noise. 
 Due to the inherent nonlinearity of the ionization process,
 a form of quantum stochastic resonance-like behavior 
 has been observed. This quantum SR leads to a dramatic
 enhancement (by several orders of magnitude) in the nonlinear ionization
 when a specific but small amount of white noise is added to the
 weak few cycle laser pulse. We have further shown the signatures of the 
 enhancement effect for different types of the laser pulses from
 a few cycles to few tens of cycles duration. Moreover, if the 
 noise amplitude is kept fixed to some level, and the peak pulse amplitude is
 varied, again a curve with a specific maximum is obtained for the enhancement parameter.
 These results suggest the existence of an optimum combination of the laser pulse and noise,
 if one is interested in optimizing the relative ionization enhancement.
 The same effect is also achieved if one uses realizable
 broadband chaotic light instead of white noise.
 We emphasize that the effect is robust with respect to a range of
 experimentally accessible parameters such as the pulse duration. 

 The enhancement mechanism is analyzed in the frequency domain 
 by measuring the frequency-resolved gain profile of the atom under a strong laser
 pulse, employing a pump-probe type of setting. 
 The gain providing frequencies are significantly modified from the unperturbed 
 atomic resonances, suggesting the non-resonant nature of the noise absorption. 
 However, if we introduce a relative time-delay between the laser pulse and noise
 the enhancement is still present, provided the noise acts first on the atom. In this case,
 the useful frequencies in the noise spectrum are at the atomic resonances.

 Finally, analogous effects are also expected in other systems
 provided the following three conditions are full-filled:
 (i) The system has a single-well finite binding potential with multiple energy levels,
 (ii) it can be subjected to a (nonresonant) coherent optical driving, 
 and (iii) it can be subjected to an incoherent perturbation. 
 Since these conditions are sufficiently general, other systems (SQUIDs, molecules
 etc) might also display similar features \cite{squid, Chelk}.

\begin{acknowledgments}  
 We thank A. Kenfack, W. Peijie, N. Singh, A. Buchleitner, and
 P. H\"anggi for fruitful discussions.
\end{acknowledgments} 
  
 \section*{APPENDIX}
  \renewcommand{\theequation}{A-\arabic{equation}}
  \setcounter{equation}{0}  

 In this appendix we briefly present an algorithm for the numerical simulation
 of the time-dependent stochastic Schr\"odinger equation, Eq.~(6) in the text, 
 where the Hamiltonian is given by Eq.~(1). 
 The properties of the white Gaussian noise $\xi(t)$ are defined
 in Eq.~(4)-(5).
 Our basic approach is to use the split-operator fast-Fourier 
 transform (SOFFT) method due to Feit and Fleck 
 (which is well known for the deterministic case) \cite{QMbook}, and adapt
 it to the case when the Hamiltonian
 contains an additional stochastic term $\xi(t)$.

 Recalling briefly that if there were no random term, the solution of 
 Eq.~(6) can be obtained by defining the standard propagator $U(t_0, t)$, 
 which when applied on initial state wavefunction $\vert \Psi_0 \rangle$
 propagates it forward in time. The usual approach to compute the solution of the 
 Schr\"odinger equation is to discretize the total propagation time into N
 small steps of equal intervals $\Delta t$. The resulting exact
 short-time propagator can be written as,
 \begin{equation}            
 U_\xi (\Delta t) = \exp\left( -i\int_{t}^{t+\Delta t} (H_{det}(x,t) + x \xi(t)) dt\right). 
 \label{eqn:A1}
 \end{equation}
 Here, $H_{det}(x,t) = \hat{p}^2/2 + V(x) + x F(t) $, is the deterministic atomic Hamiltonian
 including the laser-atom interaction term.
 In order to incorporate the white noise term within the framework
 of SOFFT method, one can rewrite the propagator as,
 \begin{equation}            
 U_\xi (\Delta t) = U (\Delta t) \exp\left( -i\int_{t}^{t+\Delta t} x \xi(t) dt\right),
 \label{eqn:A=2}
 \end{equation}
 where, $U (\Delta t) = \exp\left( -i H_{det}(x,t) \Delta t \right)$, denotes the
 deterministic part of the propagator.
 The stochastic integral in the exponential can be interpreted in the 
 Stratonovitch sense \cite{book} using the 
 properties of the white Gaussian noise as follows,
 \begin{equation}   
 \int_{t}^{t+\Delta t} \xi(t) dt = \sqrt {2D \Delta t } X_t,
 \label{eqn:A3}
 \end{equation}
 where $X_t$ is a random number having Gaussian distribution and of
 unit variance. This makes the propagator a stochastic operator.
 Note that even in the presence of the noise term the operator is 
 unitary, i.e., it preserves the norm of the wavefunction.

 One can approximate the exact propagator given by Eq.~(A-2) following a three-step
 splitting leading to the following expression,
  \begin{eqnarray}  
 U_\xi (\Delta t) =  \exp\left( -i \frac{p^2}{2} \Delta t \right) \nonumber \\
 \exp\left( -i V_\xi (x, t) \Delta t) \right)  
 \exp\left( -i \frac{p^2}{2} \Delta t \right),
 \label{eqn:A=2}
 \end{eqnarray}
 with, $ V_\xi (x, t) = V(x) + x F(t) + \sqrt {2D /\Delta t} X_t$.
 Note that the effect of noise is simulated by inserting at every time
 step a random number $X_t$ whose statistical properties are described
 above. The right hand side of Eq.~(A-4) is thus equivalent to
 the free propagation over a half time increment $\Delta t / 2$, 
 a random phase change from the action of potential $V_\xi (x, t)$ over the whole
 time $\Delta t$, and an additional free particle propagation
 over $\Delta t / 2$.

 This operator splitting is correct up to second order in the time step
 $\Delta t$ for the noise-free part, but due to the stochastic integration 
 it is accurate up to only first-order for the stochastic part. 
 In the actual calculation $\Delta t$ is chosen sufficiently small, 
 such that a further reduction in its value does not
 alter the accuracy of the physical results. 
 For a given realization of the random number sequence
 entering in the propagator via $X_t$, one generates a quantum ``trajectory'' 
 for the wavefunction. To extract the physical observable, 
 an ensemble average of the desired quantity over a large
 number of noise realizations is needed.
 Other simulation parameters such as the grid size
 and the grid resolutions should be taken as described 
 in the literature \cite{SOper}.

\end{document}